\documentclass[conference]{IEEEtran}
\usepackage{cite,graphicx,amsmath,amssymb}
\usepackage{subfigure}
\usepackage{citesort}
\usepackage{fancyhdr}
\usepackage{mdwmath}
\usepackage{mdwtab}
\usepackage{balance}
\usepackage{xcolor}
\usepackage{bm}
\usepackage{cite,graphicx,amsmath,amssymb}
\usepackage{subfigure}
\usepackage{citesort}
\usepackage{fancyhdr}
\usepackage{mdwmath}
\usepackage{mdwtab}
\usepackage{balance}
\usepackage{xcolor}
\usepackage{bm}
\usepackage{amssymb}
\usepackage{amsmath}
\usepackage{cite}
\usepackage{url}
\usepackage{xcolor}
\usepackage{cite,graphicx,amsmath,amssymb}
\usepackage{subfigure}
\usepackage{citesort}
\usepackage{fancyhdr}
\usepackage{mdwmath}
\usepackage{mdwtab}
\usepackage{caption}
\usepackage{amsthm}
\usepackage{algorithm}
\usepackage{algorithmic}
\usepackage{multirow}
\usepackage{flafter}
\usepackage{graphicx}
\usepackage{subfigure}
\usepackage{algorithm}
\usepackage{algorithmic}
\usepackage{epstopdf}

\newtheorem{remark}{Remark}
\usepackage{algorithm}
\usepackage{algorithmic}
\usepackage{multirow}
\usepackage{flafter}
\usepackage[
top    = 1.78cm,
bottom = 1.0in,
left   = 0.64 in,
right  = 0.64 in]{geometry}

\begin{document}

\title{NOMA in UAV-aided cellular offloading: A machine learning approach}
%
\author{
\IEEEauthorblockN{  Ruikang~Zhong\IEEEauthorrefmark{1}, Xiao~Liu\IEEEauthorrefmark{1}, Yuanwei~Liu\IEEEauthorrefmark{1}, and Yue~Chen\IEEEauthorrefmark{1} }

\IEEEauthorblockA{
\IEEEauthorrefmark{1} Queen Mary University of London, London, UK\\
E-mail: \{r.zhong; x.liu; yuanwei.liu; yue.chen\}@qmul.ac.uk
 } }

\maketitle
\begin{abstract}

A novel framework is proposed for cellular offloading with the aid of multiple unmanned aerial vehicles (UAVs), while non-orthogonal multiple access (NOMA) technique is employed at each UAV to further improve the spectrum efficiency of the wireless network. The optimization problem of joint three-dimensional (3D) trajectory design and power allocation is formulated for maximizing the throughput. In an effort to solve this pertinent dynamic problem, a K-means based clustering algorithm is first adopted for periodically partitioning users. Afterward, a mutual deep Q-network (MDQN)  algorithm is proposed to jointly determine the optimal 3D trajectory and power allocation of UAVs. In contrast to the conventional deep Q-network (DQN) algorithm, the MDQN algorithm enables the experience of multi-agent to be input into a shared neural network to shorten the training time with the assistance of state abstraction. Numerical results demonstrate that: 1) the proposed MDQN algorithm has a faster convergence rate than the conventional DQN algorithm in the multi-agent case; 2) The achievable sum rate of the NOMA enhanced UAV network is $23\%$ superior to the case of orthogonal multiple access (OMA); 3) By designing the optimal 3D trajectory of UAVs with the aid of the MDON algorithm, the sum rate of the network enjoys ${142\%}$ and ${56\%}$ gains than that of invoking the circular trajectory and the 2D trajectory, respectively.

\end{abstract}

\section{Introduction}

Owing to the flexible mobility, on-demand deployment, as well as their ability to establish a high probability of line-of-sight (LoS) wireless propagation \cite{UAVov.Zeng}, unmanned aerial vehicles (UAVs) have been invoked as aerial base stations for complementing terrestrial cellular networks in diverse scenarios. On the one hand, UAV-aided wireless networks are practical to be invoked as a backup when the terrestrial cellular networks which rely on ground base stations (GBSs) are paralyzed by natural disasters \cite{Disasters.Zhao}. In these scenarios, UAVs can be employed to replace terrestrial infrastructures for forming temporary communication networks to realize information transfer and disaster relief. On the other hand, UAVs can also be invoked in cellular network offloading scenarios for enhancing connectivity, throughput and coverage of the terrestrial networks \cite{Edge.Cheng}.

Since the deployment of the UAV fleet is acknowledged as a potential scheme to reduce the access congestion and improve the quality of service (QoS) \cite{JPA.Liu}, a number of related research contributions are proposed recently. In \cite{UAVCellularMW}, the authors pointed out that the UAVs can be connected with satellites to provide further connectivity for users who suffered from congested cellular networks. The authors of \cite{Edge.Cheng} optimized the trajectory of a single UAV to serve users who are distributed at the edge of cellular networks. The sum rate of these users was maximized by iteratively optimizing the user scheduling and the UAV trajectory. In \cite{Cyclical.Lyu}, the UAV was designed to fly around the GBS following a circular trajectory and users can alternately get access to the UAV when the UAV is flying over them. The authors of \cite{JPA.Liu} applied a convex optimization approach to find out the optimum hover position and power allocation for a non-orthogonal multiple access (NOMA) enhanced UAV. In \cite{NOMAoff.Kim}, the authors also adopted a circular deployment similar to \cite{Cyclical.Lyu} but employed a NOMA scheme to simultaneously provide service to users who are near and far from the UAV.

Reinforcement learning (RL) has been successfully invoked in the UAV-aided wireless networks in virtue of its capacity on solving complex, dynamic and non-convex problems \cite{liu2020} \cite{chen2019artificial}. In \cite{TD.Xiao}, an effective Q-learning paradigm was proposed for determining the optimal positions of multiple UAVs to serve ground users. To enlarge the limited state space of the Q-learning model, a combination of Q-learning and neural network (NN), namely deep Q-network (DQN) was proposed \cite{MLOV.MAO}.   Moreover, recently, the authors of \cite{4ML.UA} introduced the application of various RL algorithms in the UAV relay networks for solving resource management problems, such as multi-armed bandit learning and actor-critic learning.

Although the aforementioned literature already paved a foundation of solving challenges in the UAV-aided cellular offloading and the NOMA-enhanced UAV, the dynamic environment derived from the movement of ground mobile users was ignored in the previous research contributions \cite{chen2017caching} and the circular trajectory is not likely to be the optimal solution for the non-ideal user distribution. In order to remedy these research deficiencies, we propose the following new contributions: 1) We propose a NOMA-enhanced UAV-aided cellular offloading framework, in which multi-UAV are deployed in 3-D space to complement terrestrial infrastructures. Build on the proposed system model, we formulate the sum rate maximization problem by jointly optimizing the dynamic trajectory of multi-UAV and power allocation policy based on the channel state information of users.  2) We propose a two-step approach to solve the formulated problem. We firstly invoke the upper bounded K-mans algorithm to periodically determine user clusters. Based on the identified user association, a multi-agent  a mutual deep Q-network (MDQN) algorithm is proposed to jointly optimize UAVs' 3-D trajectory and power allocation policy to maximize the total throughput. The trajectory derived from the proposed MDQN algorithm not only enables UAVs to establish the desired channel condition with users, but also enables each agent to strive to reduce the interference.

\section{System Model}

\subsection{System Description}

Let us consider an outdoor down-link user-intensive scenario with a central GBS and a number of moving users. In order to provide further connectivity for the overloaded cellular, we propose a multi-UAV-aided cellular offloading framework as a feasible solution, where each UAV is equipped with a single antenna and employs NOMA technique. We denote the user set of GBS served users as $m \in \mathbb{M}=\{1,2,3...M\}$, and the UAV served user set can be denoted as $k \in \mathbb{K}=\{1,2,3...K\}$ and $\mathbb{M}\bigcap\mathbb{K}=\varnothing$. The users served by UAVs are allocated to $U$ cells, namely user association, where $u \in \mathbb{U}=\{1,2,3...U\}$.  Each user in cell $u$ is only served by UAV $u$, and users in cell $u$ will be clustered into several NOMA clusters. UAVs are assumed to utilize the same frequency band but different band with GBS since the GBS has tremendous transmitting power compared with the UAV. Without loss of generality, in this paper we assume that one user cluster is associated with each UAV, and in practice, multiple orthogonal resource blocks could be employed by UAVs to serve multiple user clusters.

\subsection{Mobility Models}

In this paper, two kinds of user mobility models are invoked, namely random roaming model and directional walking model.  The moving direction and speed of random roaming users are completely random in any discrete time slot $t$. Its moving angle $\theta$ and speed conform $V_u$ to the uniform distribution ${\theta\thicksim U(0,2\pi)}$ and ${V_u \thicksim U(0,V_{max})}$. Directional random walking users' movement is the vectorial sum of two vectors, a direction vector $\overrightarrow{D_d}$ with fixed direction $\theta=\Theta$, $|\overrightarrow{D_d}|= 4/5 \cdot V_{max}$ and a random vector $\overrightarrow{D_r}$, ${\theta\thicksim U(0,2\pi)}$ and $\thicksim U(0,1/5 \cdot V_{max})$. At the initial time slot of the offloading service, offloaded users are divided into several clusters which equal to the number of UAVs according to the users' spatial location. As sparked by \textbf{Remark \ref{mobility}}, it is necessary for UAVs to check the users' location and re-cluster the users after a period of time $T_r$.

\begin{remark}
\label{mobility}

Since users are roaming continuously, the initial deployments of UAVs and user clustering would no longer be optimal at a certain moment, which motivates the re-clustering of users. Re-clustering users in the service area may not necessarily increase the sum data rate but it is a necessary condition for maintaining the optimal data rate.


\end{remark}

\subsection{Propagation Model}

The channel model between each UAV and the associated users is provided by the 3GPP specifications Release 15 \cite{3gpp.36.777}. The path loss ${L_{{\text{LoS/NLoS}}}}$ between user $k$ and UAV $u$ can be expressed as \eqref{lk}, where $h_u(t)$ represents the flight altitude of UAV $u$, $f_c$ represents the carrier frequency, and the 3-D distance between UAV $u$ and user $k$ at time $t$ is denoted as ${d_{{k}}^{u}}(t)$ that

\begin{figure*}
\begin{align}\label{lk}
{{L_{\text{LoS/NLoS}}}(t) = \left\{ {\begin{array}{*{20}{c}}
  {30.9 + \left( {22.25 - 0.5{{\log }_{10}}{h_u}(t)} \right){{\log }_{10}}d_k^u(t) + 20{{\log }_{10}}{f_c},{\kern 1pt} {\kern 1pt} {\kern 1pt} {\kern 1pt} {\kern 1pt} {\text{if}}{\kern 1pt} {\kern 1pt} {\text{LoS}}{\kern 1pt} {\kern 1pt} {\kern 1pt} {\kern 1pt} {\text{link}},} \\
  {\max \left\{ {L_{{\text{LoS}}},32.4 + \left( {43.2 - 7.6{{\log }_{10}}{h_u}(t)} \right){{\log }_{10}}d_k^u(t) + 20{{\log }_{10}}f_c} \right\},{\kern 1pt} {\kern 1pt} {\kern 1pt} {\kern 1pt} {\kern 1pt} {\text{if}}{\kern 1pt} {\kern 1pt} {\text{NLoS}}{\kern 1pt} {\kern 1pt} {\kern 1pt} {\kern 1pt} {\text{link}},}
\end{array}} \right.}
\end{align}
\end{figure*}

\begin{align}\label{dt}
{{d_{{k}}^{u}}(t) = \sqrt {{h_u}^2(t) + {{\left[ {{x_u}(t) - {x_{{k}}^{u}}(t)} \right]}^2} + {{\left[ {{y_u}(t) - {y_{{k}}^{u}}(t)} \right]}^2}} }.
\end{align}

The probability of LoS is denoted as $P_{LoS}$ and described in \eqref{plos} at the top of the next page, where ${d_0} = \max [294.05\cdot{\log _{10}}{h_u}(t) - 432.94, 18]$, while ${p_1} = 233.98\cdot{\log _{10}}{h_u}(t) - 0.95$. Logically, the probability of None-LoS channel is ${P_{{\text{NLoS}}}} = 1 - {P_{{\text{LoS}}}}$.  Therefore, the mean path loss between UAV $u$ and user $k$ can be calculated by \eqref{avgloss}


\begin{figure*}
\begin{align}\label{plos}
{{P_{{\text{LoS}}}} = \left\{ {\begin{array}{*{20}{c}}
  {1,}&{if{\kern 1pt} {\kern 1pt} {\kern 1pt} \sqrt {{{\left( {d_k^u(t)} \right)}^2} - {{\left( {{h_u}(t)} \right)}^2}}  \leqslant {d_0},} \\
  {\frac{{{d_0}}}{{\sqrt {{{\left( {d_k^u(t)} \right)}^2} - {{\left( {{h_u}(t)} \right)}^2}} }} + \exp\left \{\frac{-\sqrt{\left( {d_k^u(t)} \right)^{2}-\left( {{h_u}(t)} \right)^{2}}}{p_1} +\frac{d_0}{p_1} \right \},}&{if{\kern 1pt} {\kern 1pt} {\kern 1pt} \sqrt {{{\left( {d_k^u(t)} \right)}^2} - {{\left( {{h_u}(t)} \right)}^2}}  > {d_0},}
\end{array}} \right.}
\end{align}
\end{figure*}

\begin{align}\label{avgloss}
L_k^u(t)=P_{\text{LoS}} \cdot L_{\text{LoS}} + P_{\text{NLoS}} \cdot L_{\text{NLoS}}.
\end{align}

With the considering of small scale fading, the channel gain from the UAV $u$ to user $k$ can be calculated as
\begin{align}\label{gt}
{g_k^u(t) = {H_k^u}(t) \cdot {10^{{{ - {L_k}(t)} \mathord{\left/
 {\vphantom {{ - {L_k^u}(t)} {10}}} \right.
 \kern-\nulldelimiterspace} {10}}}}},
\end{align}
where ${H_k^u}(t)$ represents the fading coefficient\cite{offloading.Lyu} between UAV $u$ and user $k$ .


\subsection{Signal Model}

Denote ${v_{u,k}}$ as the serving indicator. ${v_{u,k}}=1$ represents that the UAV $u$ is serving the user $k$, ${v_{u,k}}=0$ if otherwise. Thus, the superposition transmitting signal $x^u(t)$ of UAV $u$ can be expressed as~\cite{cui.signal}:

\begin{align}\label{xnu}
{x^u(t) = \sum\limits_{k = 1}^K {{v_{u,k}}(t)\sqrt {P_{k}^u(t)} } x_{k}^u(t) },
\end{align}
where $x_{k}^u(t)$ is the transmitting signal from UAV $u$ to user $k$, $P_{k}^u(t)$ denotes the allocated power of user $k$. As a consequence of Equation \eqref{gt} and \eqref{xnu}, the received signals at user $k$ is

\begin{align}\label{ynuk}
{y_{k}^u(t) = g_{k}^u(t)x^u(t) + {I_\text{inter}}_{k}^u(t) + {I_\text{intra}}_{k}^u(t) + \sigma _{k}^u(t) },
\end{align}
where $\sigma _{k}^u(t)$ represents the additive white Gaussian noise (AWGN). ${I_\text{inter}}_{k}^u(t)$ is the accumulative inter-cluster interference to user $k$ from other UAVs except UAV $u$ and ${I_\text{intra}}_{k}^u(t)$ represents intra-cluster interference.

The composition of ${I_\text{inter}}_{k}^u(t)$ can be expressed as


\begin{align}\label{inuk}
{{I_\text{inter}}_{k}^u(t) = \sum\limits_{s = 1,s \ne u}^U {g_{k}^s(t)\sqrt {P^s(t)} x^s(t)} },
\end{align}
where $g_{k}^s(t)$ denotes channel gain between UAV $s\neq u$ and user $k$, $P^s(t)$ represents the total power consumption of UAV $s\neq u$, which is

\begin{align}\label{pns}
{P^s(t) = \sum\limits_{k = 1}^K {{v_{u,k}}(t)P_{k}^s(t)} }.
\end{align}

The precondition of determining ${I_\text{intra}}_{k}^u(t)$ is to find out the optimal decoding order to guarantee the successful successive interference cancellation (SIC) since NOMA protocol invokes SIC to remove some of the intra-cluster interference at the receiver side \cite{SIC}. In this case, a dynamic decoding order has to be considered owing to the fact that the channel gain and inter-cluster interference of each user is always changing by the movement. The auxiliary term $G_{k}^u(t)$ shown in \eqref{gnku} is interjected as a criterion for determining the decoding order, and $G_{k}^u(t)$ can be regarded as the equivalent channel gain.

\begin{align}\label{gnku}
{G_{k}^u(t) = \frac{{{v_{u,k}}(t)g_{k}^u(t)}}{{\sum\nolimits_{s = 1,s \ne u}^T {g_{k}^s(t)P^s(t) + {{\sigma _{k}^u(t)} ^2}} }}}.
\end{align}


Consider a NOMA cluster with user $j$ and user $k$ associated with UAV $u$, and their equivalent channel gains can be noted as $G_{k}^u(t), G_{j}^u(t)$, respectively. Then the condition of user $k$ to remove the signal of user $j$ by SIC is $G_{k}^u(t) \ge G_{j}^u(t)$, which can be derived from \cite{cui.signal}. This inequality suggests that SIC is supposed to be implemented at the receiver with stronger equivalent channel gains. Extending the above principle to a NOMA cluster $u$ with $K^u$ users, we can find out an equivalent channel gain based decoding order, noted as $G_{\pi (1)}^u(t) \le G_{\pi (2)}^u(t) \le  \cdots  \le G_{\pi (K^u)}^u(t)$, where $\pi (k)$ denotes the decoding order of user $k$. According to the SIC principle, the user $\pi(k)$ decodes and successively subtracts the signals for all the $\pi (k-1)$ users, and then decode the desired signal. With this principle, since the signals for $\pi (k-1)$ users are removed, the intra-cluster interference ${I_\text{intra}}_{\pi(k)}$ and desired signal for user $\pi(k)$ be calculated as
\begin{align}\label{Iintra}
{{I_\text{intra}}_{\pi(k)} = \sum\limits_{i=k+1}^{K^u}{v_{u,\pi (i)}}(t)g_{\pi (i)}^u(t)P_{\pi (i)}^u(t)x_{\pi (i)}^u(t)}.
\end{align}
\begin{align}\label{Desired}
{S_{\pi(k)} = {v_{u,\pi (k)}}(t)g_{\pi (k)}^u(t)P_{\pi (k)}^u(t)x_{\pi (k)}^u(t)}.
\end{align}

Build on Equation \eqref{inuk} \eqref{Iintra} and \eqref{Desired}, the signal-to-interference-and-noise ratio (SINR) for the user $\pi(k)$ is given by \eqref{rnpiku}.

\begin{figure*}\label{rnpiku}
\begin{align}\label{rnpiku}
{\gamma _{\pi (k)}^u(t) = \frac{{{v_{u,\pi (i)}}(t)g_{\pi (k)}^u(t)P_{\pi (k)}^u(t)}}{{\sum\nolimits_{i = k + 1}^{ K^u } {{v_{u,\pi (i)}}(t)g_{\pi (i)}^u(t)P_{\pi (i)}^u(t) + \sum\nolimits_{s = 1,s \ne u}^U {g_{k}^s(t)P^s(t) + {{\sigma _{k}^u(t)} ^2}} } }}}.
\end{align}
\end{figure*}

Then the data rate of user $k$ connected with UAV $u$ can be calculated as

\begin{align}
\mathcal{R}_{\pi (k)}^u(t) = {B}\log 2\left( {1 + \gamma _{\pi (k)}^u(t)} \right),
\end{align}
where $B$ represents bandwidth of UAV $u$. Hence, the sum data rate at time $t$ can be calculated as
\begin{align}\label{Rnpiku}
{{\mathcal{R}(t)} = \sum\limits_{u = 1}^U \sum\limits_{k = 1}^K{{\mathcal{R}_{\pi (k)}^u(t)} } }.
\end{align}

Therefore, the throughput during the serving period is

\begin{align}\label{Ravg}
\mathcal{R} = \sum_{t=0}^{T}\mathcal{R}(t).
\end{align}

\section{Problem Formulation}

Intending to maximize the total throughput, we optimize the trajectory and power allocation policy of UAVs, subject to the maximum power constraint, spacial constraints, and the QoS constraint. The problem is formulated in \eqref{OPP}. $H = \{h_u(t),0\leq u\leq U, 0\leq t\leq T\}$ represents the positions of UAVs, and the velocity of UAVs is assumed as fixed. The transmitting power of each UAV is $P_u$, the power allocation policy is denoted as $P = \{p_k(t), 0\leq t\leq T, k \in {\mathbb K}\}$. Finally, the serving indicator $V = \{v_{u,k}(t), t = T_r, u\in {\mathbb U}, k\in {\mathbb K}\}$ is used to represent user associations. Hence, the optimization problem can be formulated as



\begin{subequations}
\begin{align}\label{OPP}
\max_{\mathbf{H,V,P}} \quad &\mathcal{R} = \sum_{t=0}^{T}\mathcal{R}(t), \\
\textrm{s.t.}
&{h_{\min }} \le {h_u(t)} \le {h_{\max }},\forall u,\forall t ,\notag \\
&{x_{\min }} \le {x_u(t)} \le {x_{\max }},\forall u,\forall t ,\notag \\
&{y_{\min }} \le {y_u(t)} \le {y_{\max }},\forall u,\forall t \label{OPPB},\\
&{\sum_{u=1}^{N} v_{u,k}=1 }, \label{OPPD}\\
&\sum\limits_{k \in {\mathbb K}} { {{v_{u,\pi (k)}}(t)P_{k}^u \le {P_u}} }, \forall t,\forall u,\forall k,\label{OPPE}\\
&G_{\pi (k)}^u \ge G_{\pi (j)}^u,k > j,\forall (k,j),\forall t,\forall u,\label{OPPF}\\
&R_k(t) \geq R_{\text{QoS}}, \forall k, \forall t, \label{OPPG}
\end{align}
\end{subequations}
where \eqref{OPPB} indicates the constraints for 3-D position of the UAV, which has to be in the airspace above chosen cellular within achievable height range to avoid the collision between UAVs or interfere other communication equipments outside the offloading cellular.  Constraint \eqref{OPPD} ensures each user $u\in \mathbb U$ only be served by one UAV. Constraint \eqref{OPPE} denotes the transmitting power constraint to guarantee the power consumption of each UAV never beyond the upper transmitting power bound. Constraint \eqref{OPPF} represents the decoding order for successful SIC.  Constraint \eqref{OPPG} formulates the rate constraint in terms of fairness of users. Since the problem category of \eqref{OPP} was proved to be NP-hard in \cite{NP.HARD}, and the formulated problem is with highly dynamic due to the movement of UAVs and users, it is challenging for the conventional convex-optimization algorithms to solve. Thus, the RL-based algorithm, which can interact with the environment and learn from its own exprences, is invoked in this paper.

\section{Proposed Reinforcement Learning Scheme}
This section introduces the proposed solution which contains two aspects, the K-means based user clustering and the joint optimization for trajectory and power allocation via MDQN.
As aforementioned, the K-means algorithm is employed to cluster users based on their spatial location at $t \in T_r$.  Additionally, since the conventional K-means algorithm cannot guarantee the uniformity of clustering, an appropriate uniformity improvement is applied when any cluster has the number of users out of the load ability of the UAV.

We propose a multi-agent MDQN algorithm to jointly optimize the UAV trajectory and power allocation. Multiple UAVs are considered as independent agents to choose actions, but multiple UAVs are permitted to connect with the same NN during the training process with the assistance of state abstraction. In the MDQN model, agents need to connect to the NN accordingly.
In this paradigm, although the experience of each agent is different, it can be reorganized into a standard form and then these experiences can be used to train a mutual NN. It can also be considered that the standardized experience of each agent can also be indirectly obtained by other agents via the shared NN. Thus, the training time is compressed and the lengthiness training problem of the conventional DQN paradigm is alleviated. It is worth to note that the MDQN paradigm only requires data exchange during the training process.  The detailed algorithm flow has been listed in \textbf{Algorithm \ref{DQN TDPA}}.

\begin{algorithm}
\caption{MDQN algorithm for deployments and power allocation}
\label{DQN TDPA}
\begin{algorithmic}[1]

		\FOR{each episode}
        \STATE Initialize initial positions of UAVs and users
        \STATE Initialize the evaluation network $w_e$ and the target network with random parameter $w_t$
        \STATE Update $\epsilon$ in action policy
        \FOR{each step $ t_0 \leq t \leq t_0 + T_r$}
		\FOR{each UAV}
        \STATE Calculate $G_k^u, k\in \mathbb{K}$
        \STATE Generate state abstraction array $S$
        \STATE Choose $A$ according to action policy and $Q(S,A,w_e)$
        \STATE Take action $A$, observe $R$ and $S'$
        \STATE Store $e = (S,A,R,S')$
        \STATE Sample random pair of $e$ from memory

        \STATE Calculate target $y = R + \beta \max Q(S',A',w_t)$
        \STATE Train parameter $w_e$ with a gradient descent step $(y - Q(S,A,w_e))^2$
        \IF{update = true}
        \STATE $w_t \leftarrow w_e$
        \ENDIF

        \STATE $S \leftarrow S'$
        \ENDFOR
        \STATE Users move
		\ENDFOR
        \ENDFOR
\end{algorithmic}
\end{algorithm}

\subsection{State abstraction}

Since the MDQN model needs to calculate the Q value of actions according to the input state information $S$, which is formed by positions of UAVs and user channel gain in this model.
In order for multiple UAVs to share the NN, the state information from each UAV have to be abstracted and shuffled into a standard array before the state information is entered into NN. The shuffling are illustrated in ~\eqref{S}, that the UAV currently connected to the neural network needs to latch its input neurons. By the feat of this design, the neural network approximates the logical relationship between the interferer and the victim and this logic is universal for all UAV with equivalent equipment.

Moreover, since $L_{u}$ and $g_{k}^u$ have different dimensions and excessively divergent magnitude, in order for the MDQN algorithm to efficiently process these mixed data, scalarization and scaling is suggested to be taken. The input state array $S$ can be expressed as

\begin{align}\label{S}
S = \{L_{u}(t),L_{s}(t),g_{k}^u(t),g_{k}^s(t)\}, u,s \in \mathbb{U}, s\neq u, k \in \mathbb{K},
\end{align}
where $L_{u}(t)$ is denote the 3D coordinate of the connecting agent and $L_{s}(t)$ denote coordinates of other agents, which are considered as sources of inter-cluster interferences. Analogously, $g_{k}^u(t)$ and $g_{k}^s(t)$ represent the channel gain of associated users and the channel gain of users associated with other UAVs, respectively.

\begin{remark}\label{Expshare}

State abstraction makes it possible for multiple agents to jointly train an NN. Compared to the approach that multi-agent train NN independently, the proposed approach can significantly increase the convergence rate. The three-phase in state abstraction, shuffling, scalarization and scaling are essential, otherwise, NN may not be able to converge.

\end{remark}

\subsection{Action Space}

The action space contains two subsets, UAV movement actions and power allocation policies for the next step. All UAVs have the same following action space:

\begin{itemize}
\item Movement action space: UAV is authorized to choose an action from seven flight actions, \{horizontal left, horizontal right, horizontal forward, horizontal backward, vertical upward, vertical downward, hover\}. Corresponding to \eqref{OPPB}.
\item Power allocation action space: Since the MDQN model outputs discrete actions, the power distribution for each user is preset to multiple gears ${P_1,P_2 \ldots P_p}$. The agent will select and maintain a power gear for each associated user until the next action.
\end{itemize}

\subsection{Action Policy}

An efficient $\epsilon-greedy$ action policy with a decreasing $\epsilon$ is adopted in training. This policy makes the agent have the probability of $\epsilon$ to choose the exploration (random action), and the probability of $1-\epsilon$ to choose the exploitation (optimal action). Mathematically, it can be expressed as
\begin{equation}
A=
\begin{cases}
random\ action, &  \epsilon, \\
{argmax}_A Q(S,A,w_e),&  1-\epsilon.
\end{cases}
\end{equation}

\subsection{Reward Function}
As mentioned in equation \eqref{OPP}, the objective function is maximizing the total throughput under the condition of guaranteeing the fairness \eqref{OPPG}, so the reward function is designed as
\begin{align}\label{Reward}
R =\frac{\mathcal{R}(t)}{2^\lambda},
\end{align}
where $R(t)$ denotes the data rate and $\lambda$ is the penalty coefficient. The penalty coefficient increases when the agent chooses a route that violates the QoS requirement. $\lambda$ stops raising when increasing it cannot reduce the number of steps that do not meet the QoS requirements.


\section{Numerical Results}

This section provides numerical results to validate the effectiveness of the proposed approaches and evaluate the gain of each component in the proposed approaches. In the simulation, users are randomly distributed in the service area and 3 UAVs are deployed near the boundary of the cellular with a height of 100 meters at the initial time. The employed neural network is with 3 layers and a 40-nodes hidden layer. The activation function is rectified linear units and mean squared error is chosen as the loss function. The Adam optimizer is applied for training the NN.  The $\epsilon$ for greedy action policy is set to linear decreasing from 0.9 to 0.

\begin{figure}[t!] 
    \begin{center}
\includegraphics[width=3.15in]{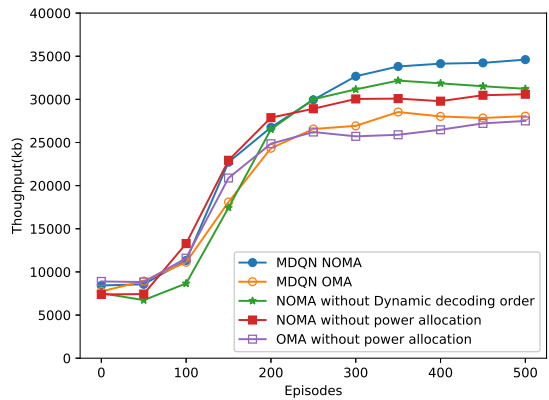} 
\caption{Throughput improvements of dynamic decoding order and DQN power allocation } 
\label{Fig.Power-Decodingorder} 
    \end{center}
\end{figure}

Fig.\ref{Fig.Power-Decodingorder} displays the throughput versus training episodes number and the curves demonstrate the convergence of the proposed multi-agent MDQN model. The throughput is significantly improved by the participation of NOMA compared with the OMA case. Meanwhile, it also figures out the contribution of power allocation and dynamic decoding order on the throughput in both NOMA and OMA cases. It can be observed that when NOMA is invoked, the dynamic decoding order and the power allocation derived from the proposed MDQN algorithm achieve gains of approximately $12\%$ and $14\%$, respectively.

\begin{figure}[H] 
    \begin{center}
\includegraphics[width=3.15in]{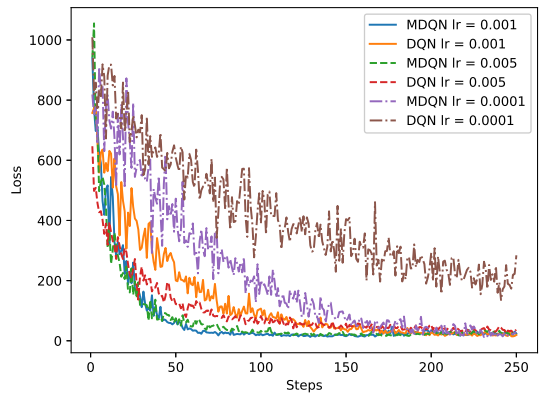} 
\caption{DQN Loss vs training steps} 
\label{Fig.Loss} 
    \end{center}
\end{figure}

\begin{figure}[htb] 
    \begin{center}
\includegraphics[width=3.15in]{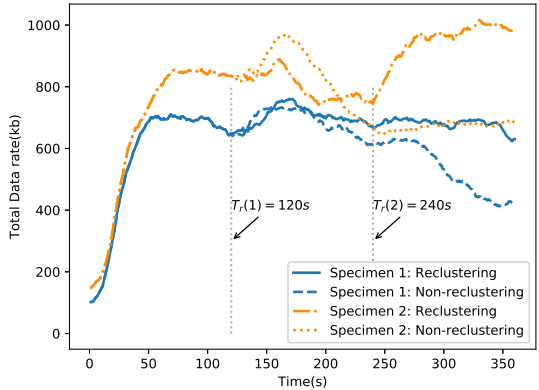} 
\caption{Data rate in test episode with/without re-clustering} 
\label{Fig.re-cls} 
    \end{center}
\end{figure}

\begin{figure}[htbp] 
    \begin{center}
\includegraphics[width=3.15in]{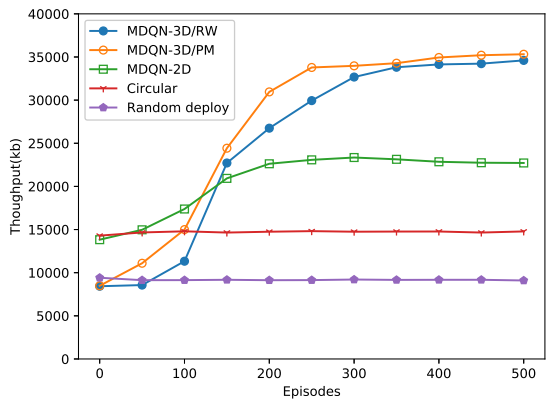} 
\caption{Throughput vs training episodes for different trajectory design scheme} 
\label{Fig.Deployment} 
    \end{center}
\end{figure}

Fig. \ref{Fig.Loss} compares the convergence rate of the MDQN and conventional DQN algorithm by plotting the loss. It can be observed that the proposed MDQN paradigm has a higher training efficiency compared with the conventional independent agent mode. In this simulation, three UAVs connect to one NN via state abstraction as expounded in \textbf{Remark \ref{Expshare}}. As a consequence, it can be seen from the 3 pairs of curves, the number of training steps required by the DQN algorithm is approximately three times of the MDQN algorithm.

Fig. \ref{Fig.re-cls} shows the data rate of two specimens of both considering and without considering re-clustering in the test episode to reveal the role and value of re-clustering. In these simulations, three UAVs are employed, and the users follow the directional movement. The same model and parameters are set up in the two shown specimens, but the users have different initial distributions and directions of movement. In both specimens,  after a period of time, the data rate of users without re-clustering receives a sustained decrease that does not appear in the re-clustered case, which can be ascribed to the lack of re-clustering since other conditions are exactly the same. This phenomenon suggests that in long-term service, re-clustering is beneficial to the data rate and it also provides the evidence for the insights in \textbf{Remark \ref{mobility}}.

Fig. \ref{Fig.Deployment} compares the trajectory derived from the proposed algorithm with the benchmarks derived from the existing literature. We test the proposed method by invoking two mentioned user mobility models RW(random walk) and PM(Purposeful movement) to prove its universality and the benchmarks are only simulated with random roaming users.  The 3-D trajectories are capable to achieve significant advantages over the 2-D trajectory and the circular trajectory. Compared to chaotic deployment, the circular trajectory has a better performance but inferior to all MDQN-derived trajectories.

\section{Conclusions}

This paper was undertaken to design an effective paradigm for employing NOMA enhanced UAVs to assist terrestrial base stations and evaluated the performance of the proposed RL algorithm. The user cluster was determined by the K-means algorithm, 3-D deployments and power allocation were jointly optimized by the proposed MDQN algorithm to maximize the total data rate of fleet-served users.
Our simulation evaluated the performance of the proposed approach from multiple dimensions, including the convergence, trajectory, and multiple access schemes through the numerical results. These results proved the superiority of the NOMA framework and the proposed MDQN paradigm possesses better convergence than the conventional DQN paradigm.

\bibliographystyle{IEEEtran}
\bibliography{NOMA_UAV_ML}

\end{document}